\documentclass[12pt]{amsart}
\input{xy}
\xyoption{all}
\usepackage{amssymb}
\usepackage[dvips]{graphicx}
\usepackage{color}
\DeclareGraphicsExtensions{.bmp,.png,.pdf,.jpg}

\newtheorem{theorem}{Theorem}[section]

\newtheorem{definition}[theorem]{Definition}

\numberwithin{equation}{section}

\newcommand{\ra}{\rightarrow}

\def\umono{\ar@{_{(}->}[u]}
\def\uumono{\ar@{_{(}->}[uu]}

\def\lmono{\ar@{_{(}->}[l]}
\def\llmono{\ar@{_{(}->}[ll]}

\newcommand{\F}{{\mathcal F}}

\def\phi{\varphi}
\def\Phi{\varPhi}

\def\RR{\Bbb R}

\def\notdivides{\not\kern 2.2pt\bigm |}

\title{Homogeneity test for functional data}
\author{Ram\'on Flores, Rosa Lillo and Juan Romo}

\date{\today}
\begin{document}

\maketitle

\begin{center}
{\small
\begin{tabular}{l} \bf Ram\'on Flores
\\ Department of Mathematics, Universidad Aut\'onoma de Madrid,  Spain
\\ e-mail: \url{ramon.flores@uam.es}
\\[2mm]
\bf Rosa Lillo
\\ Department of Statistics, Universidad Carlos III de Madrid, Spain
\\ e-mail: \url{lillo@est-econ.uc3m.es}
\\[2mm]
\bf Juan Romo
\\ Department of Statistics, Universidad Carlos III de Madrid, Spain
\\  e-mail: \url{romo@est-econ.uc3m.es}
\end{tabular}}
\end{center}

\vspace{2cm}

\begin{abstract}

In the context of functional data analysis, we propose new two sample tests for homogeneity. Based on
some well-known depth measures, we construct four different
statistics in order to measure distance between the two samples.
A simulation study is performed to check the efficiency
of the tests when confronted with shape and magnitude
perturbation. Finally, we apply these tools to measure the
homogeneity in some samples of real data, obtaining good results using this new method.

\end{abstract}

\maketitle

\section{Introduction}





In the recent years functional data analysis (FDA) has become a very active domain of research in
Statistics, because of its own interest and also for its
applications in a number of contexts such as medical science, biology,
chemistry and social sciences. In essence, the objects of study in
FDA are real functions which are assumed to be generated by means
of a stochastic process. The functions are observed in a certain
number of fixed points or time instants, but instead of being
treated as multivariate data, they are smoothed using appropriate
tools. Nevertheless, a number of techniques of multivariate data have
been adapted or generalized to the FDA context. The main
references in this field are Ramsay-Silverman \cite{Ramsay05} and
Ferraty-Vieu \cite{Ferraty06}.

In this paper, we address the problem of homogeneity between
samples of functions; that is, given two samples of curves, we
need to decide whether these two samples have been produced or
not by the same process so that they have equal probability
distributions. This problem has been recently considered. 
Our test fits in the framework of the rank test of L\'opez-Pintado and Romo \cite{Lopez09} (based in turn on \cite{Liu93}) to establish the homogeneity of
two functional samples. Other authors have confronted related problems, but more from the point of view of comparing operators rather than testing homogeneity. Benko et al. \cite{Benko09}
present methods for testing
equality of means between functional data that rely, respectively,
on bootstrap and asymptotic procedures. Horv\'ath-Kokoszka (\cite{Horvath12} and \cite{Horvath13}) also describe tests to compare the equality of
covariance operators.  Cluster
algorithms have also been proposed in \cite{Abraham03}, for
example. Finally, a different point of view is developed by
Cuevas-Febrero-Fraiman in \cite{Cuevas04}, where an F-test for analysis of variance
based on functional distances was proposed. In a similar way, the
approach we take in this paper is related to distances between the
two functional samples which are based on depth measures, and it is important to remark that we intend to test homogeneity rather than means or covariances.

 Consider an interval $T\subset \RR$, and a finite sample $\F=\{x_1\ldots x_n\}$ of real functions defined over the interval.
 We will always assume that the functions lie in $C^1(T)$.
 The concepts of distance between samples that we introduce in this paper will be based on the \emph{statistical depth}, a concept originating in the statistical analysis of multivariate data and then extended to functional data. In our context, a \emph{depth functional} with respect to the sample $\F$ will be a functional $d:C^1(T)\ra \RR$, whose value should depend in a certain way on the sample $\F$ and also on a depth measure defined a priori. In this way, the value of $d$ over the function will constitute a measure of how deep the function $f$ ``inside" the sample $\F$ is. By means of these functionals, we construct four families of statistics which are shown to be useful to decide if two samples of functions are homogeneous or heterogeneous.
 In order to understand the behaviour of the measures with respect to differences of magnitude and shape in the samples, we
have tested our methods
 on several samples of simulated functions.
 Moreover, we discuss homogeneity in some real contexts, such as Ramsay height data, the tecator sample and the mitochondrial data MCO. See the last section for details.

The structure of the paper is as follows. In Section 2 we
present the concept of depth, introduce the notion of depth with
regard to a sample and describe our statistics. Section 3 reviews
the measures of depth that are used in order to construct the
different homogeneity tests. Section 4 is devoted to the analysis
of some samples of simulated data, whereas in the last section we
perform the test for the real data examples.


\section{Distances between functional samples}

In the growing field of FDA, where functions are data, a crucial
general goal is to define concepts which mimic or transport the
usual notions in multivariate data analysis. The depth of
functions was defined -in any of its versions- in this sense, as
a generalization to this context of a notion of centrality, with
the deepest function of a certain sample being an adequate definition of
the ``median" of the data.

 In order to introduce intuitive statistics that indicate in some way distance between two
samples, we propose the definition of depth of a sample with
respect to another one. Given a certain measure of functional depth $d$, and given a
sample of functions $\mathcal{F}$ and another function $g$ not
necessarily in $\mathcal{F}$, we denote by $d_{\mathcal{F}}(g)$
the depth of $g$ with regard to the sample $\mathcal{F}\cup \{
g\}$. We define henceforth the notion of the deepest function of a
sample with respect to another:

\begin{definition}

Let $\F$ and $\mathcal{G}$ be two finite samples of continuous functions defined in an interval $T$. The \emph{deepest function} of $\mathcal{G}$ with regard to $\F$ is the function $g$ of the sample $\mathcal{G}$ which maximizes $d_{\mathcal{F}}(g)$ among $g\in \mathcal{G}$. We will denote this function by $\mathfrak{D}_{\F}(\mathcal{G})$, or simply $\mathfrak{D}(\mathcal{G})$ if the base sample $\mathcal{F}$ is understood. If there is more than one function in $\mathcal{G}$ for which the depth is reached, we can choose any of them as $\mathfrak{D}_{\F}(\mathcal{G})$, or else we can consider the whole set as the \emph{deepest subsample
 of} $\mathcal{G}$\emph{ with regard to} $\F$.

\end{definition}

Observe that if the samples $\F$ and $\mathcal{G}$ are large enough, the probability of finding two functions of $\mathcal{G}$ which maximize the depth decreases, so usually we can talk about the deepest function.

The definition of the deepest function of $\mathcal{G}$ with
respect to $\F$ is addressed to propose a solution to the problem
of homogeneity in the context of functional data. Given two or
more samples of functions, we say that the functions are
homogeneous if they come from the
same experiment, and then have equal probability distributions. In
our framework, explicitly determining the distributions is
usually a very difficult problem, so we are forced to design
different strategies to test homogeneity. We offer hence a
different approach to the problem, by using the depth measures to
perform an analysis which, by nature, may include the distance
between functions, their magnitude and their shape.

Our starting point are two samples of functional data, $\mathcal{F}$
and $\mathcal{G}$. The functions
$\mathfrak{D}_{\mathcal{F}}(\mathcal{G})$ or
$\mathfrak{D}_{\mathcal{G}}(\mathcal{F})$ may supply interesting
information about homogeneity. In this sense the concept of deepest
function can be used in different ways. In the following, we propose
several possible statistics which depend on the notion of deepest
function and allow us to undertake the analysis of homogeneity.


 We define the first statistic $\mathbf{P_1}$ as
$$\mathbf{P_1}(\mathcal{F},\mathcal{G})=d_{\mathcal{F}}\mathfrak{D}_{\mathcal{G}}\mathcal{G}.$$
Probably this is the more natural approach to the homogeneity
problem, since roughly speaking, the function
$\mathfrak{D}_{\mathcal{G}}\mathcal{G}$ is the most representative element of
the experiment which produces the sample $\mathcal{G}$. Hence, it
is reasonable to compute how deep this estimator is with respect
to $\mathcal{F}$. The greater this depth, the less likely
the two samples come from different experiments.

The second statistic is defined as a variation of the previous
one:

$$\mathbf{P_2}(\mathcal{F},\mathcal{G})=|\mathbf{P_1}(\mathcal{F},\mathcal{G})-\mathbf{P_1}(\mathcal{F},\mathcal{F})|.$$

 This definition may be considered a kind of normalization of the previous one. It could happen that the nature of the experiment which originates the sample $\mathcal{F}$ makes impossible for any datum of the experiment to reach the value 1 (for example, if the experiment produces two well-defined ``bands" of functions, or if some deep functions cross each other in close points). In this case, $\mathbf{P_1}(\mathcal{F},\mathcal{F})$ would give a good estimation of the maximum of these depths, and the difference $|\mathbf{P_1}(\mathcal{F},\mathcal{G})-\mathbf{P_1}(\mathcal{F},\mathcal{F})|$ would be more informative than the value $\mathbf{P_1}(\mathcal{F},\mathcal{G})$ alone; see the computations section for interesting questions about this issue. In this case, the samples are likely to come from the same experiment as the statistic gets closer to zero.  It would probably be equivalent to consider the quotient instead of the absolute value of the difference.

A different approach is given by the statistic
$$\mathbf{P_3}(\mathcal{F},\mathcal{G})=d_{\mathcal{F}}(\mathfrak{D}_{\mathcal{F}}\mathcal{G}),$$
which identifies the $\mathcal{F}$-depth of the deepest function
of $\mathcal{G}$ with respect to $\F$. This is the function of the
sample $\mathcal{G}$ which is more likely to come from the
experiment that generates the sample $\F$, and then it is relevant
from the point of view of the classification. In particular,
$|\mathbf{P_3}(\mathcal{F},\mathcal{F})|=\mathbf{P_1}(\mathcal{F},\mathcal{F})$
for any sample $\mathcal{F}$.

 Observe that the function
$\mathfrak{D}_{\mathcal{F}}\mathcal{G}$ could not be a good
estimator for the result of the experiment that generates
$\mathcal{F}$. Then, if we intend to use it for the classification
of experiments, it would also be interesting to produce a measure
that controls simultaneously the $\F$-depth and the
$\mathcal{G}$-depth of $\mathfrak{D}_{\mathcal{G}}\mathcal{G}$.
One possible option is to define a measure in $[0,1]\times [0,1]$
whose values are the $\F$-depth and the $\mathcal{G}$-depth of
$\mathfrak{D}_{\mathcal{F}}\mathcal{G}$; in this context, the
first number would be the measure of the depth itself, while the
second would be interpreted as a control number of how sharp
the measure is. However, this approach is bivariant, so we propose
instead an alternative univariant version that avoids that
disadvantage and captures essentially the same information:

$$\mathbf{P_4}(\mathcal{F},\mathcal{G})=|\mathbf{P_3}(\mathcal{F},\mathcal{G})-\mathbf{P_1}(\mathcal{F},\mathcal{F})|
|\mathbf{P_3}(\mathcal{F},\mathcal{G})-\mathbf{P_1}(\mathcal{G},\mathcal{G})|.$$

    The greater this number, the less likely the two samples come from the same experiment.



Once the statistics are defined, we propose the following method
 for testing the null hypothesis of equality of distributions of the two functional samples.
We use a bootstrap approach to test the null hypothesis of homogeneity.

\begin{enumerate}

\item[1.] Select a functional depth measure $d_{\mathcal{F}}$ and
a statistic $\mathbf{P}=\mathbf{P}_i$ for some $i\in \{1,2,3,4\}$,
which will depend on the previous concrete choice. In this paper,
to deal with $d_{\mathcal{F}}$ we will use Fraiman-Muniz depth,
$h$-modal depth, random depth $RPD$, band depth $BD$ and modified
band depth $mBD$, but there are other possible choices for the
depth measure.

\item[2.] Now consider the samples $\mathcal{F}$ and
$\mathcal{G}$, and propose as a null hypothesis $H_0$ that
$\mathcal{F}$ and $\mathcal{G}$ come from the same experiment. We
perform then a hypothesis test to reject (or not) $H_0$.

\item[3.] Define the sample $\mathcal{H}$ as the union
$\mathcal{F}\cup \mathcal{G}$, and obtain $N$ bootstrap  samples
of $\mathcal{H}$ of size $|\mathcal{H}|$. For any
$1\leq j\leq N$, let $\mathcal{S}_j$ be the corresponding sample,
denote by $\mathcal{S}^1_j$ the sample of the first
$|\mathcal{F}|$ functions and by $\mathcal{S}^2_j$ the sample of
the last $|\mathcal{G}|$. Then compute
$P_j=\mathbf{P}(\mathcal{S}_j^1,\mathcal{S}_j^2)$.

\item[4.] For an appropriate size $\alpha$, compute a
confidence interval for the values $P_j$.

\item[5.]  The null hypothesis will be rejected if and only if the
functional $\mathbf{P}(\F,\mathcal{G})$ does not belong to the
interval, and in this case we will assume that the groups are
not homogeneous.

\end{enumerate}

The nature of the computations suggests using one-sided confidence intervals. To obtain the critical value at $95\%$ of confidence, we trim five percent of the data in the appropriate side of the interval: in measures $\mathbf{P}_1$ and $\mathbf{P}_3$ there should be smaller values, while in the normalized one values will be larger.

Below we present our results with both simulated data and real
data, but we first review the functional depths we use.

\section{Functional depths}

The concept of depth in the context of functional data analysis
generalizes the same notion for multivariate data. While the multivariate
measures are mainly addressed to explore a certain centrality of a
point in some real vector space, the different nature of the
functional data forces the statistics to consider other features
of the functions involved, such as the shape of the functions or the
amount of time they spend in a certain range of real numbers. In
this sense, we have chosen different depth measures which in turn
explore different features of the functions inside the samples.
We start with the pioneering work of Fraiman-Muniz, whose goal is
to measure how much time every function is deep inside the sample.

\textbf{Fraiman-Muniz depth}. Consider a sample of curves $\{x_1(t),\ldots x_n(t)\}$ defined on
the interval $[0,1]$. Denote by $I(-)$ the indicator function and consider,
for every $i\in 1\ldots n$, the function:

$$F_{n,t}(x_i(t))=\frac{1}{n}\sum_{k=1}^nI(x_k(t)\leq x_i(t)),$$ and also the univariate depth $$D_n(x_i(t))=1-|\frac{1}{2}-F_{n,t}(x_i(t))|.$$ Then, the Fraiman-Muniz depth of the function $x_i(t)$ is defined in \cite{Fraiman01} as the integral:

$$FM(x_i(t))=\int_0^1D_n(x_i(t))dt.$$

\textbf{h-modal depth}. This measure was first defined by Cuevas
et al.  \cite{Cuevas06} and is addressed to identify the
functional mode of the sample. Consider again a sample of curves
$\{x_1(t),\ldots x_n(t)\}$, select a value $h$ which should be
interpreted as a bandwidth, and also consider a kernel function
defined on the real positive numbers. Then the h-modal depth of
the function $x_i(t)$ with respect to $K$ and $h$ is defined as:

$$hD_n(x_i,h)=\sum_{k=1}^n\frac{K(\|x_i-x_k\|)}{h}.$$

In this paper, as recommended by the aforementioned authors, we
take the norm $L^2$, $h$ as the 15th percentile of the empirical
distribution of the norms $\|x_i-x_k\|$, and $K$ being a convenient
truncated Gaussian kernel.

\textbf{Random projection depths}. These two versions of depth
were proposed by Cuevas et al.  \cite{Cuevas07}, and combine
random projections of the functions of the sample in different
directions with a bivariate data depth which is used to order the
corresponding results. More precisely, given a sample of
functions $\{x_1(t),\ldots x_n(t)\}$ and $\nu$, a realization of a
stochastic process whose values are random directions, we define
the projection of $x_i$ along the direction $\nu$ as

$$T_{i,\nu}=\int_0^1\nu(t)x_i(t)dt,$$ and analogously, $$T'_{i,\nu}=\int_0^1\nu(t)x'_i(t)dt$$ considering the derivatives instead of the trajectories of the function. If we select a bivariate data depth $D$ and assume $P$ realizations of $V$, we may define the following two versions of the random projection depth:

$$RPD1(x_i)=1/P\sum_{p=1}^PD(T_{i,\nu},T_{i,\nu}),$$ which takes into account only the trajectories of the functions, and $$RPD2(x_i)=1/P\sum_{p=1}^PD(T_{i,\nu},T'_{i,\nu})$$ which considers the functions and their derivatives.

On this note, the role of $D$ to compute depths will be played by the bivariate version of $h$-modal depth. Moreover, we checked that the results obtained in our contexts using RPD1 and RPD2 were similar, and as the second one was computationally harder, in this paper we use only the first version, which we will denote simply by RPD.

\textbf{Band depth}. In \cite{Lopez09}, L\'opez-Pintado and Romo
define two different versions of a new depth of essentially
geometric nature. It is based on the concept of band, understood
as a portion of the plane that is delimited by the sample of
curves. More precisely, fix the sample $\mathcal{F}$, and given a
continuous function defined in $T$, denote by $G(x)$ the graph of $x$. Then, for
every $j$ such that $2\leq j\leq n$, the \emph{n-th band depth }is
defined by:

$$BD_n^{(j)}(x)={\binom{n}{j}}^{-1}\sum_{1\leq {i_1}\leq {i_2}\leq \ldots \leq {i_j}\leq n}I\{G(x)\subseteq B(x_{{i_1}},x_{i_2},\ldots ,x_{i_j})\}.$$

Here $x_{{i_1}},x_{i_2},\ldots ,x_{i_j}$ are functions in the sample and $B(x_{{i_1}},x_{i_2},\ldots ,x_{i_j})$ is:

$$B_j(x)=B_j(x;x_{i_1},\ldots,{i_j})=\{(t,y)\in T\times\mathbb{R}:\textrm{min}_{k=1_1,\ldots i_j}x_k(t)\leq y\leq \textrm{max}_{k=1_1,\ldots i_j}x_k(t)\leq x(t)\}.$$

Here $I$ stands, as usual, for the indicator function. Note that $BD_n^{(j)}(x)$ measures the proportion of $j$-uplas $(x_{{i_1}},x_{i_2},\ldots ,x_{i_j})$ in $\F$ such that $x$ belongs to the band determined by them.

Next we review the global band depth, that compiles all the previous measures.

Given a sample $\F$ as above and a value $J$ such that $2\leq J\leq n$, the band depth of a function $x$ is defined as $$BD_{n,J}(x)=\sum_{j=2}^J BD_n^{(j)}(x).$$

Of course, from an analytic point of view, the most logical choice
for $J$ is $n$, so we collect all the posible information given by
the curves in the sample $\F$. However, if $|\F|$ is big, the
depth can become computationally intractable. The authors
prove that the value is quite stable in $J$, so in this paper we
will use $J=2$. In this case, the depth depends generally on
non-degenerate bands.


The authors also define a modified version of the band depth, by considering bands in the interval $T$, instead of bands in the plane:

$$A_j(x)=A_j(x;x_{i_1},\ldots,{i_j})=\{t\in T:\textrm{min}_{k=1_1,\ldots i_j}x_k(t)\leq x(t)\leq \textrm{max}_{k=1_1,\ldots i_j}x_k(t)\leq x(t)\}.$$
 Now the authors consider a Lebesgue measure $\lambda$ on the interval (usually the standard one), and define as in the previous case:

$$mBD_n^{(j)}(x)={\binom{n}{j}}^{-1}\sum_{1\leq {i_1}\leq {i_2}\leq \ldots \leq {i_j}\leq n}\frac{\lambda(A_j(x))}{\lambda(T)},$$ again with $2\neq j\neq n$. Now the definition of the modified band depth is analogous to the previous one:

$$mBD_{n,J}(x)=\sum_{j=2}^J BD_n^{(j)}(x),$$
for $2\leq J\leq n$.




\section{Simulation study}

In order to describe the characteristics and features of our
procedures, we perform a simulation study using the four different
statistics defined in Section 2   and the five depth measures
defined in the previous section: Fraiman-Muniz, h-modal, random
measure, band depth and modified band-depth. We consider six
functional populations defined in $[0,1]$, which are
considered as the realizations of a stochastic process $X(-)$
which has continuous trajectories in the interval $[0,1]$.

\textbf{Sample 0}. This is the reference set,
generated by a Gaussian process $$X(t)=E(t)+e(t)$$ with mean function
$E(t)=E(X(t))=30t^{3/2}(1-t)$, and $e(t)$ is a centered Gaussian
process, whose covariance matrix is given by
$Cov(e_i,e_j)=0.3*\textrm{exp}(-\frac{|t_i-t_j|}{0.3})$.

The remaining sets are produced by perturbing the generation process in two ways. The first three suffer magnitude contamination in the mean, while the covariant matrix does not change.

\textbf{Sample 1}. This sample is generated by the Gaussian
process $X(t)=30t^{3/2}(1-t)+1+e(t)$.

\textbf{Sample 2}. In this case the contamination is smaller than
in Sample 1: $X(t)=E(X(t))=30t^{3/2}(1-t)+0.5+e(t)$.




The next samples are obtained from the reference set of Sample 0 by changing the mean function in a more drastic way, and also the covariance matrix in some of them. These changes give rise to shape contamination.

\textbf{Sample 3}. This set is generated by the Gaussian
process $X(t)=30t(1-t)^2+e(t)$, where $e(t)$ is defined in the
same way as above.

\textbf{Sample 4}. Defined as
$X(t)=30t(1-t)^2+h(t)$, where $h(t)$ is a centered Gaussian
process whose covariance matrix is given by
$Cov(e_i,e_j)=0.5*\textrm{exp}(-\frac{|t_i-t_j|}{0.2})$.

\textbf{Sample 5}. The last group combines the previous
cases, being defined by $30t^{3/2}(1-t)+h(t)$. Hence, the
perturbation here is only induced by the process $h(t)$.

The routines used to undertake the simulations were developed in
R  and are available upon request. We adopt the following
notation: for $i\in \{1,2,3,4,5{}\}$ the five sets of simulated functions will be denoted by $S_i$, and for every $k\in \{1,2,3,4\}$, the  statistic $\mathcal{P}_k$
used in the hypothesis test  will be as defined in Section 2.


\begin{figure}[h]
\centering
\includegraphics[width=1\textwidth]{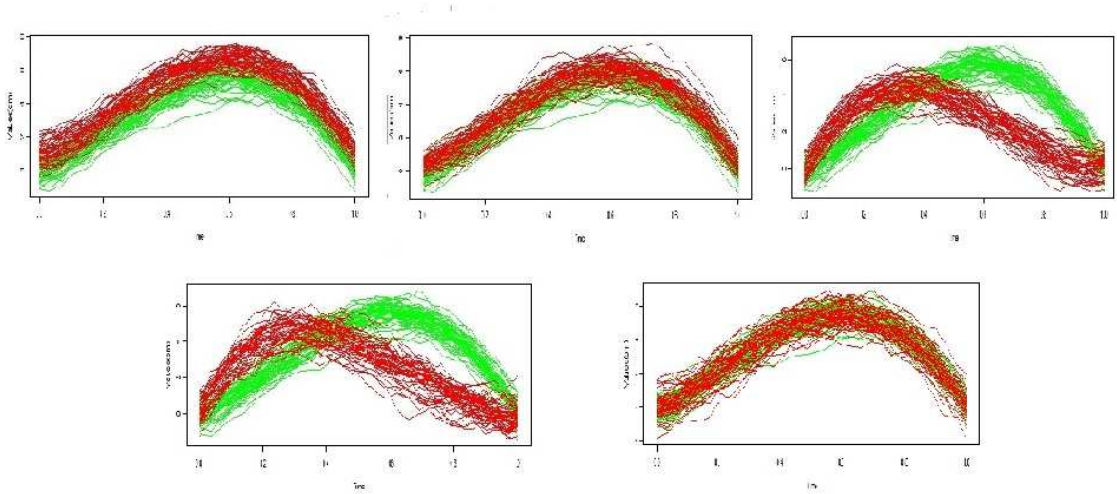}
\caption{From left to right and from up to down, the five samples. In green, the reference sample.}
\label{contexto:figura}
\end{figure}

We use the following method to test homogeneity. Select a depth
measure $d_{\mathcal{j}}$ and a statistic
$\mathbf{P}=\mathbf{P}_k$ from the list above. Now generate 50
functions with the algorithm for $S_0$ and 50 functions with the
algorithm for $S_i$, for a certain $i$. Each curve is observed in
30 equidistant points.  Now compute $\mathbf{P}_i(S_0,S_i)$. Then
consider 1000 standard bootstrap samples of size 100 of the sample
$\mathcal{H}=S_0\cup S_i$. For any $1\leq j\leq 1000$, let
$\mathcal{S}_j$ be the corresponding sample, denote by
$\mathcal{S}^1_j$ the sample of the first 50 functions and by
$\mathcal{S}^2_j$ the sample of the last 50, and compute
$P_j=\mathbf{P}(\mathcal{S}_j^1,\mathcal{S}_j^2)$. With this 1000
values we compute an one-sided confidence interval for a
confidence of 0.05. Now the null hypothesis is that $\mathcal{F}$
and $\mathcal{G}$ come from the same experiment, and we reject if
and only if $\mathbf{P}(\F,\mathcal{G})$ does not belong to the
interval $T$. Finally we repeat the whole process 100 times and
count the number of rejections. Our results are shown in the
tables, and commented below.


The results of our computations are listed in
Table \ref{MSD} with the information of
both the previous measures and  the rank tests. There we denote
respectively by FM, dmode, RPD, BD and mBD, the  Fraiman-Muniz
depth, $h$-modal depth, random projection
depth, and band depth and modified band depth. We maintain the
notation for the samples of functions which are already described
and are the target of our study. For each statistic, the table
shows the number of rejections in 100 essays, for the usual level
0.05 of confidence.


We may analyze our results from three differents point of view,
focusing respectively in the classification criteria, the depth
measures or the populations. Considering criteria, it is clear from
the data that the most accurate is $\mathbf{P_3}$ as it always
distinguishes the samples, with a perfect $100\%$ of success. Its
normalized version works also quite well, being uneffective when
combining it with h-modal depth, or when the magnitude
contamination is too small. The measure $\mathbf{P_3}$ only
presents  problems when its associated depth is BD, and same
phenomenon happens to $\mathbf{P_4}$.

From the point of view of the depth measures, it is clear all of
them work well (at least 75 rejections in almost all the cases)
except the band depth; so in case we need to use this kind of
measure, the modified version is clearly preferable. Finally it
is apparent from the simulations that the difficulties only arise
if the magnitude contamination is really small (Sample 2) or we
combine the two perturbations (Sample 5) and the measures are
powerful when confronted with other types of contaminations. Note
that, excluding BD, the measures always detect the difference for
samples $S_1$, $S_3$ and $S_4$ and
$S_5$.




\subsection{Sensitivity analysis}

We carry out a sensitivity analysis for our approach with respect to
several aspects that can be considered:

\emph{Size of the bootstrap}. In order to test the importance of
the size of the bootstrap sample, we also undertook some test
cases enlarging it to 1000 and 3000. The computation time
increased in a significant way, while there was not an apparent
change in the conclusions of our study. So we may conclude that
our statements are stable with regard to the size of the
bootstrap resampling.

\emph{Significance}. We choose the usual signification level of
0.05, but in order to check the robustness of our results, we
tested some of the data for a level of 0.025. We obtain the same
conclusions as in the 0.05 case, so we may assume that our measures
are also robust in this sense.

\emph{Symmetry}. We also check what happens if in each case, we
take the population $S_i$ as the reference sample in the hypothesis
test, and $S_0$ as the test sample. Again, the results were
similar to the ones that are shown in the paper. While a priori it
would be a good idea to take into account this symmetric values,
we check that the benefit of this strategy would be exiguous, and
at the same time the computational cost would increase
significantly.

\emph{Power test}. In order to show the performance of the
measures introduced in the paper, we have carried out a  power
test for a concrete model case. Consider the Gaussian stochastic
process $X(t)=30t^{3/2}(1-t)+e(t)+\eta$, which depends on the
parameter $\eta>0$, and consider the measure $\mathbf{P}_1$
regarding Fraiman-Muniz depth. We know from
Table \ref{MSD} that for $\eta=1$
and $\eta=0.5$ the measure separates this sample from the
reference sample in 100 out of 100 replications. After generating
another 100 replications for $\eta=0.25$, we found that the
measure detected heterogeneity in all cases. However, for
$\eta=0.1$ the sample gets really close to Sample 0 and then the
measure only discriminates in 9 out of 100 cases.

\begin{table}

\begin{tiny}

\begin{tabular}{|c|c|c|c|c|}

\hline
 & $\mathbf{P}_1$ & $\mathbf{P}_2$ & $\mathbf{P}_3$  & $\mathbf{P}_4$  \\
 \hline
FM & &  & &  \\
 \hline
  1
& 100 & 100 & 100 & 100 \\
\hline
 2 & 100 & 100 & 100 & 100 \\
\hline
 3 & 100 & 100 & 100 & 100 \\
\hline
 4 & 100 & 100 & 100 & 100 \\
 \hline
 5 & 57 & 52 & 100 & 76\\
\hline
h-modal & &  & &  \\
 \hline
  1
& 100 & 100 & 100 & 48 \\
\hline
 2 & 95 & 92 & 100 & 80 \\
\hline
 3 & 100 & 100 & 100 & 29 \\
\hline
 4 & 100 & 100 & 100 & 100 \\
 \hline
 5 & 87 & 70 & 100 & 83 \\
\hline
RPD & &  & &  \\
 \hline
  1
& 100 & 100 & 100 & 98 \\
\hline
 2 & 100 & 100 & 100 & 28 \\
\hline
 3 & 100 & 100 & 100 & 100 \\
\hline
 4 & 100 & 100 & 100 & 100 \\
 \hline
 5 & 16 & 19 & 100 & 36 \\
\hline
BD & &  & &  \\
 \hline
1 & 100 & 19 & 100 & 97
 \\
\hline
 2 & 52 & 21 & 100 & 25 \\
\hline
 3 & 100 & 32 & 100 & 93 \\
\hline
 4 & 46 & 30 & 100 & 92 \\
 \hline
 5 & 65 & 22 & 100 & 73 \\
\hline
mBD & &  & &  \\
 \hline
  1
& 100 & 100 & 100 & 76 \\
\hline
 2 & 100 & 99 & 100 & 41 \\
\hline
 3 & 100 & 100 & 100 & 100 \\
\hline
 4 & 100 & 100 & 100 & 100 \\
 \hline
 5 & 67 & 46 & 100 & 83 \\
\hline

                                 \end{tabular}
\caption{\label{MSD} Simulation results}
\end{tiny}
\end{table}

We also confronted the reference Sample 0 with itself using the homogeneity test described above for all the four statistics $\textbf{P}_i$ and the five depth measures. The proportion of rejections in each case is depicted in Table \ref{TD2}. The results show that the proportion is always smaller than 0.1, and in twelve out of twenty cases is not larger than the significance level 0.05. Therefore, the homogeneity test presents a reliable behavior with respect to the power under the null hypothesis.

\vspace{1cm}
\begin{table}
\begin{tiny}

\begin{tabular}{|c|c|c|c|c|}
\hline
\textbf{Power test 2} & $\textbf{P}_1$ & $\textbf{P}_2$ & $\textbf{P}_3$ & $\textbf{P}_4$ \\
\hline
 FM &  0.02 & 0.04 & 0.04 & 0.03 \\
 \hline
 h-modal & 0.9 & 0.04 & 0.09 & 0.04 \\
 \hline
 RPD & 0.07 & 0.05 & 0.09 & 0.02 \\
 \hline
 BD  & 0.06 & 0.05 & 0.03 & 0.06 \\
 \hline
 mBD & 0.05 & 0.06 & 0.03 & 0.07 \\
 \hline

                                 \end{tabular}
\end{tiny}
  \caption{\label{TD2} Power test for the reference sample}
\end{table}

\vspace{1cm}

Observe that in the same situation (see Table \ref{RTSD}) the rank test produces a perfect score for samples 1, 3 and 4, but it fails to prove homogeneity when the difference of magnitude is small (Sample 2) or when the shape contamination is important (Sample 5). The latter was advised early in \cite{Lopez09}.

\vspace{1cm}
\begin{table}
\begin{tiny}
\begin{tabular}{|c||c|c|c|c|c|c|}

\hline
\textbf{Rank test simulated data} & FM & h-modal & RPD  & BD  &  mBD \\
 \hline
Sample 1 & 100  & 100 & 100  & 100  & 100  \\
 \hline
Sample 2 & 65  & 41 & 57  & 49  & 44  \\
\hline
Sample 3  & 100  &  100 & 100 &  100 & 100 \\
 \hline
Sample 4  & 100 & 100 & 100  & 100  & 100  \\
 \hline
Sample 5  & 61  & 94 & 6  & 99  & 78 \\
\hline

                                 \end{tabular}

\caption{\label{RTSD} Rank test for simulated data}
\vspace{1cm}
\end{tiny}
\end{table}

To prevent disfunctions caused by outliers, it is usual to define
trimmed measures, considering a subsample of functions in $F$,
for example, $95\%$ of deeper functions. The smaller these
numbers are, the greater the probability that both series of data
come from the same experiment. We have checked the trimmed
measures in some of our previous computations, but the results
were very similar to the measures without trimming, so we offer
here the results of the latter.












\section{Real data}

In this last section, we illustrate the validity of our methods
with four different  real data sets: a) \emph{Ramsay growth curves
dataset}, which consists of the height (in cm) of 93 people
measured throughout time; b) \emph{MCO data}, where data measure
calcium content in cardiac cells of mice; c) \emph{Tecator
spectrometric data set}, which consists of 215 infrared spectra of
meat samples obtained by a Tecator IFF Analyzer, and d) the
second derivative of the spectrometric data. The results of the
rank test are included at the end of the section.

In the tables below, CV (critical value) stands for the extreme of the one-sided confidence interval of the test. Observe that for the measures $\mathbf{P}_1$ the null-hypothesis is rejected when the value of the statistic is smaller than CV, whereas in the remaining two we reject when
the value of $\mathbf{P}_n$ is larger than CV. In the corresponding columns labeled ``Rej." we specify if the null-hypothesis is rejected or not in each case.

\subsection{Ramsay data}

\begin{figure}[h]
\centering
\includegraphics[width=1\textwidth]{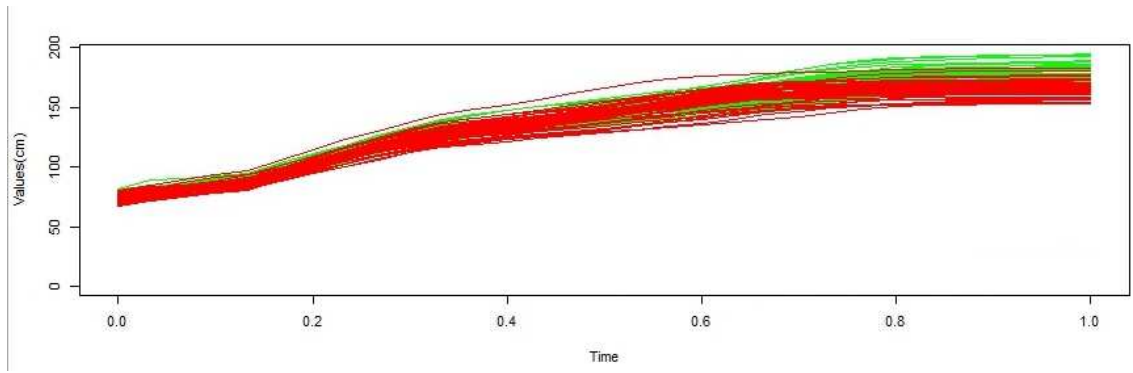}
\caption{Ramsay growth data}
\label{contexto:figura}
\end{figure}

We start our analysis of real cases with the classical growth
dataset first studied by Ferraty-Vieu in \cite{Ferraty06}, and
also analyzed more recently by L\'opez-Pintado and Romo in
\cite{Lopez09} and by Alonso, Casado and Romo in \cite{Alonso12}. The
variables are the 93 growth curves for 39 boys and 54 girls,
measured between 1 and 18 years of age, and we attempt to test the
homogeneity of samples by sex.  The results are shown in
Table \ref{RD}.

\vspace{1cm}
\begin{table}
\begin{tiny}

\begin{tabular}{|c||c|c|c||c|c|c||c|c|c||c|c|c|}
\hline
 & $\mathbf{P}_1$ & CV & Rej. & $\mathbf{P}_2$ & CV & Rej. & $\mathbf{P}_3$ & CV & Rej. & $\mathbf{P}_4$ & CV & Rej. \\
 \hline
FM & $0.879$ & 0.827 & No & $0.128$ & 0.089 & Yes & $0.893$ & 0.888 & No & $0.0007$ & 0.001 & No \\
 \hline
dmode  & $6.685$  &  6.153 & No & $3.157$   & 0.722  & Yes & $4.908$   &  6.579 & Yes & $4.389$  & 1.276 & Yes \\
\hline
RPD  & $0.224$  & 0.211  & No &  $0.088$  & 0.03 & No & $0.239$  & 0.24 & Yes & $0.00002$  & $0.00004$ & Yes \\
\hline
BD & 0.05 & 0.147 & Yes & 0.271 & 0.125 & Yes & 0.194 & 0.204 & Yes & 0.011  & 0.006 & Yes \\
\hline
mBD & 0.392 & 0.46 & Yes & 0.121 & 0.0572 & Yes & 0.497 & 0.499 & Yes & $0.00004$ & $0.0003$ & No \\
\hline
 \end{tabular}
 \end{tiny}
 \caption{\label{RD} Measures for Ramsay data}
\end{table}
\vspace{1cm}

It is obtained that a $95\%$ level of confidence, the measure
$\mathbf{P}_2$ establishes a clear difference between male and
female data for the four considered depths. Moreover, the four
statistics separate when they are combined with
band-depth and modified band depth. The ``natural" measure
$\mathbf{P}_3$ is effective in four out of five cases, and the
remaining one (when combining with Fraiman-Muniz) is very close
to being so. For these data, only $\mathbf{P}_1$ seems to be not
quite so powerful, as it separates only when combined with $BD$ and
$mBD$. Looking at the 24 outcomes of
Table \ref{RD},  we obtain 70.8
percent level of separations, which increases to 83.3 percent if we do
not take into account the measure $\mathbf{P}_1$. Observe also
that for these data the rank test only separates in half of the
cases, and in particular is ineffective for mBD. It is also
remarkable that both methods show weakness when combined with
Fraiman-Muniz depth, which
 seem not quite appropriate to confront these kind of observations.

\subsection{MCO data}

\begin{figure}[h]
\centering
\includegraphics[width=1\textwidth]{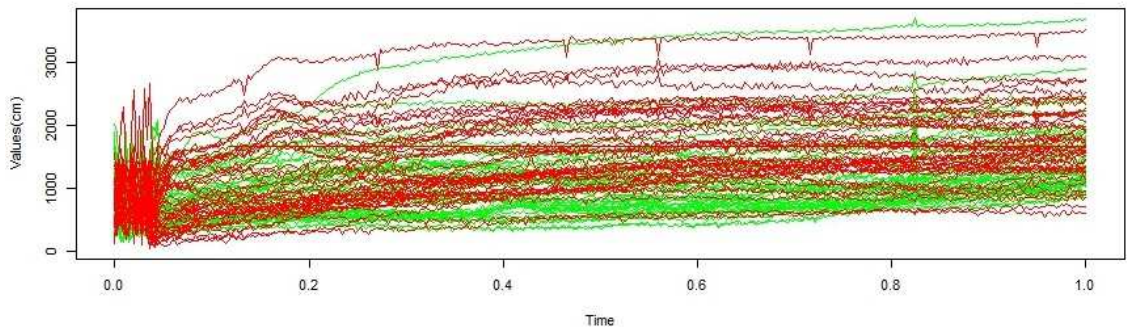}
\caption{Mitochondrial calcium data}
\label{contexto:figura}
\end{figure}

Now we apply our measures to the mitochondrial calcium overload dataset (\cite{RuizMeana03}), previously studied from a statistical point of view in \cite{Cuevas06} and \cite{Baillo11}.
The functional variable measures the level of mitochondrial calcium in mouse cardiac cells, as high levels of this element usually imply good protection of these cells in the event of ischemia process. The ultimate goal of the study is to test the power of the drug Cariporide to increase the levels of calcium in the cells. The dataset consists of control group of 45 observations and a treated group of 44. The levels of MCO are measured every ten seconds during an hour, so each function is observed in principle at 360 points; however, the data which correspond to the first three minutes are eliminated from the sample, as they show a high variability which depend on factors that are hard to control.

In Table \ref{MCD} we present the
results of our computations for the mitochondrial data MCO.

\vspace{1cm}
\begin{table}
\begin{tiny}

\begin{tabular}{|c||c|c|c||c|c|c||c|c|c||c|c|c|}

\hline
 & $\mathbf{P}_1$ & CV & Rej. & $\mathbf{P}_2$ & CV & Rej. & $\mathbf{P}_3$ & CV  & Rej. & $\mathbf{P}_4$ & CV & Rej. \\
 \hline
FM  & 0.689 & 0.814 & Yes & 0.263 & 0.127 & Yes & 0.922 & 0.894 & No & $0.0002$ & 0.002 & No \\
 \hline
dmode & 4.425 & 5.865 & Yes & 2.649 & 1.392 & Yes & 6.854 & 6.799 & Yes & 0.046 & 0.416 & No \\
\hline
RPD  & 0.227 & 0.205 & No & 0.0366 & 0.033 & Yes & 0.244 & 0.239 & No & $0.000007$ & $0.00004$ & No \\
\hline
BD & 0.047 & 0.07 & Yes & 0.096 & 0.088 & Yes & 0.078 & 0.111 & Yes & 0.008 & 0.002 & Yes \\
\hline
mBD & 0.338 & 0.449 & Yes & 0.181 & 0.077 & Yes & 0.502 & 0.498 & No & $0.00026$ & $0.00034$ & No \\
\hline

 \end{tabular}

 \end{tiny}
  \caption{\label{MCD} Measures for MCO data}
  \end{table}
\vspace{1cm}

 Different from the case of Ramsay data, we do not know a priori if the data are naturally split into two samples or not. Again the measure $\mathbf{P}_2$ offers the greatest evidence for the splitting hypothesis, as it shows heterogeneity in all the cases. The measure $\mathbf{P}_1$ also offers support to that hypothesis, as it only fails to make a difference when combining with the random depths. Measure $\mathbf{P}_3$ only rejects homogeneity in half of the cases, and $\mathbf{P}_4$ just one. From the point of view of the depth, band-depth shows again difference in all the samples.
 For these data, the rank test shows heterogeneity in two cases, when it is carried out with the $h$-modal and band depths.

\subsection{Tecator}

\begin{figure}[h]
\centering
\includegraphics[width=1\textwidth]{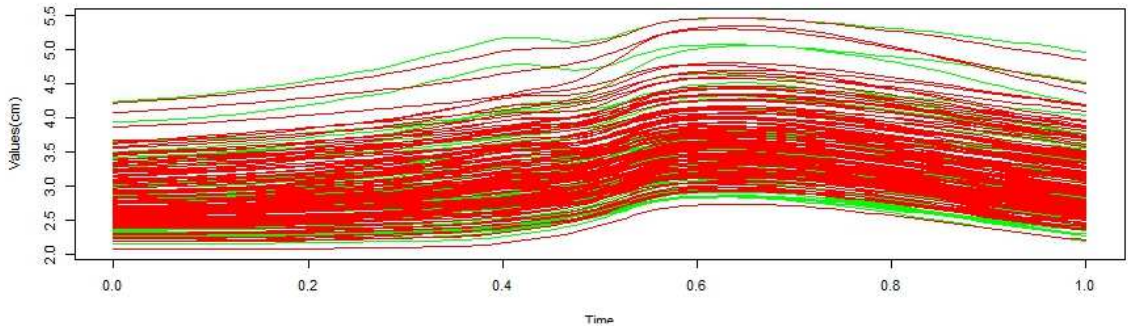}
\caption{Tecator data}
\label{contexto:figura}
\end{figure}

The tecator dataset has been intensively studied in recent
years, see for example \cite{Ferraty06}, \cite{Li08} and
\cite{Martin13}.  Tecator is a commercial name for a Infracted
Food Analyzer, which in this case is used to measure the infrared
absorbance spectrum of meat samples. These absorbances are given
as functions of the intensity of the light measured just before
and just after passing through the sample. The observations
measure the contents of moisture, protein and fat in every sample
of meat, and the goal is to separate two samples according to
their different levels of fat. The discrete observations consist
of 100 channel-absorbance spectrum for a given wavelength, which
are made continuous using a B-spline basis of order 6. The dataset
is divided into data with high fat content (77 observations) and
data with low content (the remaining 158). Following the approach of the
aforementioned papers of Ferraty-Vieu and Li-Yu, we have computed
our homogeneity measures also for the spectrometric data and for the second derivative of it.
 Recall that the discrete derivative is defined by means of the differences between subsequent points where the values for the functions are taken. The results for the first case appear in Table \ref{contexto:figura}:

\vspace{1cm}
\begin{table}

\begin{tiny}

\begin{tabular}{|c||c|c|c||c|c|c||c|c|c||c|c|c|}
\hline
 & $\mathbf{P}_1$ & CV & Rej. & $\mathbf{P}_2$ &  CV & Rej. & $\mathbf{P}_3$ & CV & Rej. & $\mathbf{P}_4$ & CV & Rej. \\
 \hline
FM & $0.946$  & 0.866   & No & $0.032$   & 0.094  & No & $0.983$   & 0.959  & No & $0.00001$  & $0.0002$ & No \\
 \hline
dmode  & $9.301$  & 7.686  & No & $0.144$   & 2.062  & No & $9.446$   & 8.91  & No & $2.008$  & 3.173 & No \\
\hline
RPD  & $0.243$  & 0.233 & No &  $0.007$  & 0.016 & No & $0.249$ &   0.248 & No  & $0.0000004$  & $0.000003$ & No \\
\hline
BD & 0.382 & 0.344 & No & 0.085 & 0.095 & No & 0.457 & 0.417 & No & $0.0002$ & 0.001 & No \\
\hline
mBD & 0.511 & 0.476 & No & 0.007 & 0.037 & No & 0.518 & 0.514 & No & 0.003 & $0.00002$ & Yes \\
\hline
\end{tabular}

 \end{tiny}
  \caption{\label{TD} Measures for tecator data}
\end{table}
\vspace{1cm}

Our computations support the widespread impression that the meat
samples of the tecator data may proceed from the same sample. As
just one out of our 24 measures is able to separate the data
(concretely $\mathbf{P}_4$ combined with modified band-depth), it
is quite likely that this is an outlier instead of a genuine
difference. Moreover, it can be seen that the critical values are
usually quite far from the extremes of the corresponding interval.

More evidence is extracted from the rank test, which shows homogeneity in the five cases, and always in a quite robust way.  The evidence then suggests that we cannot reject the hypothesis of equality between the two samples.

\subsection{Tecator second derivatives}

\begin{figure}[h]
\centering
\includegraphics[width=1\textwidth]{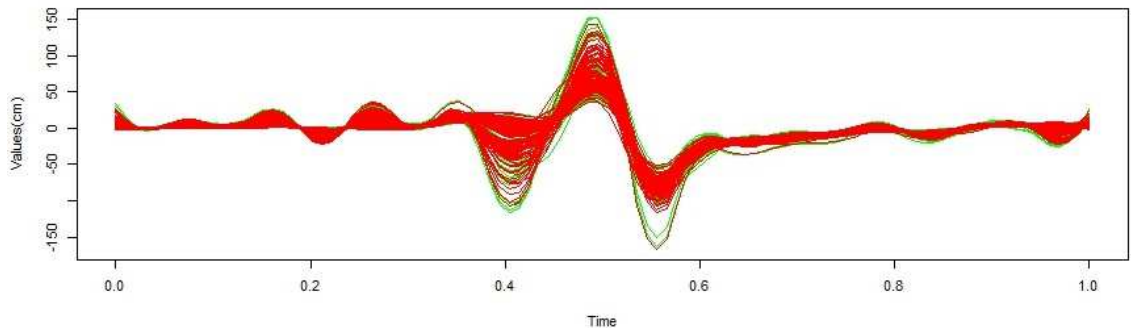}
\caption{Tecator second derivatives}
\label{contexto:figura}
\end{figure}

There is quite a lot more evidence of heterogeneity in the sample of the
second derivatives, as we may check in Table \ref{TD2}.

\vspace{1cm}

\begin{table}
\begin{tiny}

\begin{tabular}{|c||c|c|c||c|c|c||c|c|c||c|c|c|}
\hline
 & $\mathbf{P}_1$ & CV  & Rej. & $\mathbf{P}_2$ & CV & Rej. & $\mathbf{P}_3$ & CV & Rej. & $\mathbf{P}_4$ & CV & Rej. \\
 \hline
FM & $0.835$  & 0.859 & Yes & $0.091$   & 0.067 & Yes & $0.919$   & 0.887 & No & $0.0001$  & $0.0008$ & Yes \\
 \hline
dmode  & $10.581$  &  12.032 & Yes & $2'729$   & 0.722 & Yes & $13'413$   & 12.402 & No & $0.874$  & 5.663 & No \\

\hline
RPD  & $0.215$  &  0.225  & Yes &  $0.025$  & 0.017 & Yes & $0.237$  & 0.233 & No & $0.00003$  & $0.00008$ & No \\
\hline

 BD & 0.121 & 0.089 & No & 0.056 & 0.1 & No & 0.179 & 0.121 & No & $0.00006$ & 0.00437 & No \\
  \hline

 mBD & 0.448 & 0.461 & Yes & 0.054 & 0.036 & Yes & 0.498 & 0.482 & No & 0.101 & $0.0003$ & Yes \\
  \hline
 \end{tabular}

  \end{tiny}
   \caption{\label{TD2} Measures for tecator data (second derivatives)}
\end{table}
\vspace{1cm}

 As in the previous sample MCO, both $\mathbf{P}_1$ and $\mathbf{P}_2$ are able to separate, in this case four out of five cases, and again the other two measures do not seem too powerful in this case. The scheme is very similar to that case, except for the fact that band-depth gives no difference in any of the four cases. It is also remarkable that $\mathbf{P}_4$ only separates when combining with modified band-depth, just as it happens in the possible outlier case described above.

In this case the rank-test supports the hypothesis of non
homogeneity, as it is shown in all of the five observations.

\vspace{1cm}
\begin{table}
\begin{tiny}

\begin{tabular}{|c|c|c|c|c|}
\hline
\textbf{Rank test} & Ramsay & MCO & Tecator & Tecator 2 \\
\hline
 FM &  1733 & 2140 & 8427 & 7737  \\
 \hline
 h-modal & 1233 &  1625 & 8553 & 8490 \\
 \hline
 RPD &  1721 & 2051 & 8296 & 7768 \\
 \hline
 BD  &  1159 & 1482 & 8136 & 6989 \\
 \hline
 mBD &  1703 &  2140 & 8427 & 7757 \\
 \hline
 \hline
  CV & 1623.095 & 1781.395 & 7595.08 & 7595.08 \\
  \hline
                                 \end{tabular}
\end{tiny}
  \caption{\label{TD} Rank test in real data}
\end{table}

\vspace{1cm}

\section{Discussion}

In this paper, we have defined some new measures of distances
between samples of functions to solve the problem of homogeneity
in the context of functional data analysis. Combining these
measures with the depth functions defined by Fraiman-Muniz,
Cuevas-Fraiman-Muniz and L\'opez-Pintado-Romo, we propose a
hypothesis test based on the bootstrap methology and apply it to a
number of simulated and real functional data. Our measures shows
their effectiveness in detecting differences of magnitudes and
shape in some samples generated by Gaussian processes, and
moreover are able to show heterogeneity for Ramsay data,
mitochondrial data and the second derivatives tecator data. It is
significant that our methods show homogeneity in the tecator
data without differentation, a phenomenon widely dealt with in the
literature. It is also noteworthy that our method improves
the rank-test in some cases.

Once the concept of depth of a function with regard to a sample is
defined, several generalizations appear to be possible. For
example, the sample of tecator data discussed above shows that
there is information about homogeneity hidden in the derivatives
that cannot be directly extracted from the original functions.
Hence, it should be interesting to define and describe a unified
way to deal with all the depth measures and statistics used in our
work when applied at the same time to all the functions and all
their derivatives. It is likely that such a notion would be able
to show patterns in the homogeneity of the samples that could not
be deduced without differentiation. On the other hand, it would be
also interesting to define some measures that allow us to test at the
same time the homogeneity of several samples of functions. We
plan to undertake this task in subsequent work.

\end{document}